\documentclass[aps,prc,twocolumn,showpacs,floatfix,nofootinbib,superscriptaddress,amsmath,amssymb,longbibliography]{revtex4-1}

\usepackage{graphicx}
\usepackage{epsfig}
\usepackage{bm}
\usepackage{color}
\usepackage{float}
\usepackage{dcolumn}
\usepackage{multirow} 
\usepackage{hyperref}

\usepackage{amsmath}

\graphicspath{{.}{../}}

\newcommand{\beq}{\begin{equation}}
\newcommand{\eeq}{\end{equation}}
\newcommand{\beqa}{\begin{eqnarray}}
\newcommand{\eeqa}{\end{eqnarray}}

\begin{document}
\title{Microscopic optical potentials for calcium isotopes}
\thanks{This manuscript has been authored by UT-Battelle, LLC under
  Contract No. DE-AC05-00OR22725 with the U.S. Department of
  Energy. The United States Government retains and the publisher, by
  accepting the article for publication, acknowledges that the United
  States Government retains a non-exclusive, paid-up, irrevocable,
  world-wide license to publish or reproduce the published form of
  this manuscript, or allow others to do so, for United States
  Government purposes. The Department of Energy will provide public
  access to these results of federally sponsored research in
  accordance with the DOE Public Access
  Plan. (http://energy.gov/downloads/doe-public-access-plan).}

\author{J.~Rotureau}

\affiliation{NSCL/FRIB Laboratory, Michigan State University, East
  Lansing, Michigan 48824, USA}

\author{P.~Danielewicz}

\affiliation{NSCL/FRIB Laboratory, Michigan State University, East
  Lansing, Michigan 48824, USA}

\affiliation{Department of Physics and Astronomy, Michigan State
  University, East Lansing, MI 48824-1321}

\author{G.~Hagen}
\affiliation { Physics Division, Oak Ridge National Laboratory, Oak Ridge, TN 37831, USA}
\affiliation{Department of Physics and Astronomy, University of Tennessee,
Knoxville, TN 37996, USA}

\author{G.~R.~Jansen}
\affiliation { National Center for Computational Sciences, Oak Ridge National 
Laboratory, Oak Ridge, TN 37831, USA}
\affiliation { Physics Division, Oak Ridge National Laboratory, Oak Ridge, TN 37831, USA}

\author{F. M.~Nunes}
\affiliation{NSCL/FRIB Laboratory, Michigan State University, East
  Lansing, Michigan 48824, USA}
\affiliation{Department of Physics and Astronomy, Michigan State
  University, East Lansing, MI 48824-1321}

\begin{abstract}
  We construct nucleonic microscopic optical potentials by combining the Green's
  function approach with the coupled-cluster method for $\rm{^{40}Ca}$
  and $\rm{^{48}Ca}$.  For the computation of the ground-state of
  $\rm{^{40}Ca}$ and $\rm{^{48}Ca}$, we use the coupled-cluster method
  in the singles-and-doubles approximation, while for the A = $\pm 1$
  nuclei we use particle-attached/removed equation-of-motion method
  truncated at two-particle-one-hole and one-particle-two-hole
  excitations, respectively. Our calculations are based on the chiral
  nucleon-nucleon and three-nucleon interaction $\rm{NNLO_{sat}}$,
  which reproduces the charge radii of $^{40}$Ca and $^{48}$Ca, and
  the chiral nucleon-nucleon interaction $\rm{NNLO_{opt}}$.  In all
  cases considered here, we observe that the overall form of the
  neutron scattering cross section is reproduced for both
  interactions, but the imaginary part of the potential, which reflects
  the loss of flux in the elastic channel, is negligible. The latter points
  to neglected many-body correlations that would appear beyond the
  coupled-cluster truncation level considered in this work. We show
  that, by artificially increasing the parameter $\eta$ in the Green's
  function, practical results can be further improved.
\end{abstract}
\maketitle 

\section{Introduction}
Nuclear reactions are the primary experimental tool to study atomic
nuclei. With the recent progress in the development of rare-isotopes
beams (RIBs), regions of the nuclear chart far from stability, that
were previously out of reach, are now becoming accessible.  More
progress is expected, with future projects at RIB facilities, to
explore  systems far from stability \cite{fac,frib_lab,fair}.
In parallel to
the progress on the experimental side, efforts should be pursued on
the theoretical front to develop or extend reaction models to nuclei
far from stability.

It is customary in reaction theory to reduce the many-body picture to
a few-body one where only the most relevant degrees of freedom are
retained \cite {ReactionsBook}. In that case, one introduces effective
interactions, the so-called optical potentials, between the clusters
considered.  Traditionally, these interactions have been constrained
by data, particularly data on stable isotopes. Consequently, the use
of these potentials to study exotic nuclei is unreliable and has
uncontrolled uncertainties. In order to advance the field of nuclear
reactions, it is then critical to connect the effective interaction to
an underlying microscopic theory, so that
extrapolations to exotic regions are better under control, together
with rigorous assessment of uncertainties.

Realistic ab-initio nuclear
structure calculations based on nucleon-nucleon ($NN$) and
three-nucleon forces (3NFs) from chiral effective field
theory~\cite{vankolck1999,epelbaum2009,machleidt2011} have now reached
the point where reliable predictions for nuclei as heavy as
$^{100}$Sn~\cite{morris2018} can be made. This progress is due to the
development of many-body methods that scale polynomially with system's
size~\cite{mihaila2000b,dean2004,dickhoff2004,hagen2008,hagen2010b,tsukiyama2011,roth2012,hergert2013b,soma2013b,binder2013,lahde2014}
and an ever increase of computational power. On the other hand, the
ab-initio nuclear reaction community is behind in reached mass number,
precision, and accuracy. There has been a lot of effort in developing
microscopic reaction theories for light nuclei starting from realistic
$NN$ and
3NFs~\cite{nollett2007,quaglioni2008,hupin2013,elhatisari2015,mazur2015,quaglioni2012,gazit2006,efros2007,girlanda2010,marcucci2013,navratil2016},
while less so for medium-mass and heavy
nuclei \cite{hagen2012c,bacca2014b,bacca2014,idini2017,gfccpap}. With upcoming
experiments on rare isotopes in the medium- and heavy-mass region of
the nuclear chart, it is important to develop ab-initio reaction
theory that can make accurate predictions in these mass regions. It is
the aim of this paper to take the first steps towards this goal.

In this paper, we will present ab-initio calculations of
 nucleon-nucleus optical potential for the doubly magic nuclei $^{40}$Ca and $^{48}$Ca. This
work is the follow up of a previous study \cite{gfccpap}. The optical
potential (also known as the self-energy) enters the Dyson equation
together with the one-body Green's function. Assuming some
approximations for the self-energy, the standard way of obtaining the
optical potential is to iterate the non-linear Dyson equation until a
self-consistent solution is obtained. This is known as the
self-consistent Green's function
approach~\cite{dickhoff2004,barbieri2005,DOM,barbieri2017}. Our
approach differs from the self-consistent Green's function approach in
that the optical potential is obtained directly by inverting the Dyson
equation~\cite{gfccpap}. We calculate the single particle Green's
function by combining the coupled-cluster
method~\cite{kuemmel1978,bishop1991,mihaila2000b,dean2004,bartlett2007,binder2013b,hagen2014}
with the Lanczos continued fraction
method~\cite{dagotto1994,hallberg1995,haxton2005,efros2007,bacca2014b,gfccpap}
and employing a complex Berggren
basis~\cite{berggren1968,michel2002,idbetan2002,hagen2004,hagen2006}. In
this work we focus on the chiral $NN$ and $NNN$ interaction
$\rm{NNLO_{sat}}$, which has been shown to produce accurate
ground-state energies and charge radii from light to medium mass
nuclei~\cite{n2losat,hagen2015,garciaruiz2016,lapoux2016,duguet2017}.

This paper is organized as follows. In Sec.~\ref{sec_formalism}, we
 briefly revisit the formalism of the Green's function and the
coupled-cluster method along with the Berggren basis.
We start Sec.~\ref{secnumerics} by showing the convergence pattern of the optical potentials associated with the bound states in
$^{41}$Ca and $^{49}$Ca and then, present cross sections results for the neutron elastic scattering on $^{40}$Ca and $^{48}$Ca.
For comparison, we also show calculated elastic cross sections obtained with the $\rm{NNLO_{opt}}$ \cite{n2loopt} interaction (also
derived within the chiral-EFT approach) and the phenomenological
Koning-Delaroche (KD) potential \cite{KD}.
Finally, we conclude and discuss future possible applications in Sec.~\ref{conclusion}.
\section{Formalism} \label{sec_formalism}
\subsection{The single-particle Green's function} \label{spgf}
Let us consider a nucleus with A nucleons. The single-particle Green's
function for that nucleus has matrix elements
\begin{eqnarray}
G(\alpha,\beta,E)&=&\langle \Psi_{0}|a_{\alpha}\frac{1}{E-(H-E^{A}_{gs})+i \eta} a^{\dagger}_{\beta}|\Psi_0\rangle \nonumber \\
&+& \langle \Psi_{0}|a^{\dagger}_{\beta}\frac{1}{E-(E^{A}_{gs}-H)-i \eta} a_{\alpha}|\Psi_0\rangle ,
\label{gf1}
\end{eqnarray}
where $\alpha$ and $\beta$ represent single-particle states, and
$|\Psi_0\rangle$ represents the ground state of the nucleus with energy
$E^{A}_{gs}$. By definition, the parameter $\eta$ is such that in the physical limit $\eta
\rightarrow 0^+$. The operators $a^{\dagger}_\alpha$ and $a_\beta$
create and annihilate a nucleon in the single-particle state $\alpha$
and $\beta$, respectively, and their labels are shorthands for the quantum numbers
$\alpha=(n,l,j,j_z,\tau_z)$.  Here, $n,l,j,j_z,\tau_z$ label the
radial quantum number, the orbital angular momentum, the total orbital
momentum, its projection on the $z-$axis, and the isospin projection,
respectively.  The intrinsic Hamiltonian $H$ reads
\begin{eqnarray}
H=\sum^{A}_{i=1}\frac{\vec{p_i}^2}{2m}
-\frac{\vec{P}^2}{2mA} +\sum_{i<j} V_{ij} +\sum_{i<j<k} V_{ijk}, 
\label {hami}
\end{eqnarray}
with $\vec{p}_i$ the momentum of the nucleon $i$ of mass $m$ and
$\vec{P}=\sum_{i=1}^A\vec{p}_i$ the momentum associated with the
center of mass motion. The terms $V_{ij}$ and $V_{ijk}$ are $NN$ and
3NFs, respectively. It is useful to rewrite the Hamiltonian as
\begin{eqnarray}
H=\sum^{A}_{i=1}\frac{\vec{p}_i^2}{2m} \left (1-\frac{1}{A}\right)   +\sum_{i<j} \left ( V_{ij} -\frac{\vec{p}_i\vec {p}_j}{mA} \right )  +\sum_{i<j<k}V_{ijk}  ,~~~~ \label {hami2}
\end{eqnarray}
where one separates the one-body and two- (three-)body contributions.
In the following, we  work with the single-particle basis solutions
of the Hartree-Fock (HF) potential generated by $H$.
We recall here that the HF basis is a good starting point for
coupled-cluster calculations and that the HF Green's function denoted
as $G^{(0)}$ is a first order approximation to the Green's function
(\ref{gf1}).
The Green's function fulfills the Dyson equation 
\begin{eqnarray}
G(\alpha,\beta,E)&=&G^{(0)}(\alpha,\beta,E) \nonumber \\
&+&\sum_{\gamma,\delta}G^{(0)}(\alpha,\gamma,E)\Sigma^{*}(\gamma,\delta,E)G(\delta,\beta,E) .
\label{dys}
\end{eqnarray}
Here, $\Sigma^{*}(\gamma,\delta,E)$ is the self energy, which can be obtained 
from the inversion of  Eq.~(\ref{dys}):
\begin{eqnarray}
\Sigma^{*}(E)=[G^{(0)}(E)]^{-1}-G^{-1}(E) .
\end{eqnarray}
Finally, one obtains the optical potential as
\begin{eqnarray}
\Sigma'\equiv \Sigma^*+U, \label{optdef}
\end{eqnarray}
where $U$ is the HF potential. For $E\geq E^{A}_{gs}$, $\Sigma'$ in
Eq.~(\ref{optdef}) corresponds to the optical potential for the
elastic scattering from the $A$-nucleon ground
state~\cite{capuzzi,dickhoff}.  For $E\leq E^{A}_{gs}$, $\Sigma'$ has
a discrete number of solutions which correspond to the bound states in
the A+1 nucleon system.  The optical potential is non-local,
energy-dependent and complex \cite{dickhoff}; for $E\geq
E^{A}_{gs}$, its imaginary component describes, by construction, the
loss of flux due to absorption into channels other than the elastic
channel.

In this paper,  the optical potential is obtained by inverting
the Dyson equation (\ref{dys}) after a direct computation of the Green's
function (\ref{gf1}) with the coupled-cluster method \cite{hagen2014}.
In the following, we present the main steps involved in the
computation of the Green's function in our approach.

\subsection{Coupled-cluster approach for the Green's function}
In this section we briefly show how we construct the Green's function
following the coupled-cluster method.  For a more detailed account, we
refer the reader to \cite{bartlett2007,hagen2014,gfccpap}.  In
coupled-cluster theory, the ground state is represented as
\begin{eqnarray}
|\Psi_{0}\rangle=e^T|\Phi_{0}\rangle \label{cc1} , 
\end{eqnarray}
where $T$ denotes the cluster operator which gets expanded in the number of particle-hole  excitations
\begin{eqnarray}
T &=& T_1+T_2+\dots \nonumber \\
&=& \sum_{i,a}t_i^a a^{\dagger}_a a_i+\frac{1}{4}\sum_{ijab}t_{ij}^{ab}t_{ijab}a^{\dagger}_aa^{\dagger}_ba_ja_i +\ldots .
\label{t_cluster}
\end{eqnarray}
The operators $T_1$ and $T_2$ induce $1p$-$1h$ and $2p$-$2h$
excitations of the HF reference, respectively. Here, the
single-particle states $i,j,...$ refer to hole states occupied in the
reference state $|\Phi_{0}\rangle$ while $a,b,...$ denote valence
states above the reference state.  In practice, the expansion
(\ref{t_cluster}) is truncated. In the coupled-cluster with singles
and doubles (CCSD) all operators $T_i$ beyond $i=2$ are neglected.

One can show that the CCSD ground state is an eigenstate of the
similarity-transformed Hamiltonian $\overline{H}\equiv e^{-T}He^T$ in
the space of $0p$-$0h$, $1p$-$1h$, $2p$-$2h$ configurations. Note that
the transformed Hamiltonian is not Hermitian because the operator
$e^{T}$ is not unitary. As a consequence, $\overline{H}$ has left- and
right-eigenvectors which constitute a bi-orthogonal basis with the
following completeness relation
\begin{eqnarray}
\sum_{i}|\Phi_{i,R}\rangle \langle \Phi_{i,L}|=\hat{{1}},
\end{eqnarray}
where the right ground state $|\Phi_{0,R}\rangle$ is the reference
state $|\Phi_0\rangle$, while the left ground-state is given by
$\langle \Phi_{0,L} | = \langle \Phi_0 |(1 + \Lambda)$ with $\Lambda$
a linear combination of particle-hole de-excitation operators.

Using the ground state of the similarity-transformed Hamiltonian
$\overline{H}$, we can now write the coupled-cluster Green's function
$G^{CC}$ as
\begin{eqnarray}
\lefteqn{G^{CC}(\alpha,\beta,E) \equiv }\nonumber\\
&& \langle \Phi_{0,L}|\overline{a_{\alpha}}\frac{1}{E-(\overline{H}-E^{A}_{gs})+i\eta}\overline{a^{\dagger}_{\beta}}|\Phi_{0}\rangle \nonumber \\
&+&\langle \Phi_{0,L}|\overline{a^{\dagger}_{\beta}}\frac{1}{E-(E^{A}_{gs}-\overline{H})-i\eta}\overline{a_{\alpha}}|\Phi_{0}\rangle .
\label{gfcc}
\end{eqnarray}
In the expression above, $\overline{a_{\alpha}}=e^{-T}a_{\alpha}e^T$ and
$\overline{a^{\dagger}_{\beta}}=e^{-T}a^{\dagger}_{\beta}e^T$ are the
similarity-transformed annihilation and creation operators,
respectively. We note that the truncation of the cluster operator $T$ is reflected
in the expression of the coupled-cluster Green's function (\ref{gfcc})
and, if all excitations up to $Ap$-$Ah$ were taken into account in the
expansion (\ref{t_cluster}), the Green's function (\ref{gfcc}) would
be exact and identical to (\ref{gf1}).  In principle, the Green's
function could be obtained by inserting completeness relations into
(\ref{gfcc}), with the solutions of the the $A\pm 1$ systems obtained
with the particle-attached equation-of-motion (PA-EOM) and
particle-removed equation-of-motion (PR-EOM) coupled-cluster
methods~\cite{gour2006}. However, in practice, this approach is
difficult to pursue as the sum over all states also involves
eigenstates in the continuum. To bypass this issue, we use the Lanczos
continued fraction technique
\cite{dagotto1994,hallberg1995,haxton2005,efros2007,bacca2014b,gfccpap}
for the computation of the the Green's function Eq.~(\ref{gfcc}).
\subsection{Berggren basis} \label{berg} Our goal is to compute the
optical potential for elastic scattering at arbitrary
energies. However, as $\eta \rightarrow 0^+$, the coupled-cluster
Greens' function in Eq.~(\ref{gfcc}) has poles at energies
$E=(E^{A+1}_i-E^{A}_{gs})$ (with $E^{A+1}_i$ the eigenvalues of the
A+1 system), which make the numerical calculation unstable.

In order to bypass this issue, we consider, as we did in
\cite{gfccpap}, an analytic continuation of the Green's function in
the complex-energy plane.  This is achieved by working in the complex
Berggren basis (generated by the HF potential), which includes bound-,
resonant, and discretized non-resonant continuum
states~\cite{berggren1968,michel2002,idbetan2002,hagen2004,hagen2006}. In
our previous work we considered only $NN$ interactions, while in this
work we also consider 3NFs, and the transformation of the Hamiltonian
to the Berggren basis is thus much more involved. In order to obtain
the Berggren HF basis, and transform the Hamiltonian with 3NFs to this
basis, we follow the numerically efficient procedure outlined in
Ref.~\cite{ccemerg}. As a consequence, the many-body spectrum for the $A+1$ ($A-1$) systems
obtained with the PA-EOM (PR-EOM) is composed of bound, resonant and
complex-continuum states {\it i.e.} the poles of the analytically
continued Green's function have either a negative real or complex
energy. In that case, the Green's function matrix elements for $E\geq
0$ smoothly converge to a finite value as $\eta \rightarrow 0^+$.

In order to fulfill the Berggren completeness \cite{berggren1968}, the
complex-continuum single-particle states must be located along a
contour $L^+$ in the fourth quadrant of the complex momentum plane,
below the resonant single-particle states.  According to the Cauchy
theorem, the form of the contour $L^+$ is not important,
as long as  all resonant states lie between the contour and the real
momentum axis.  The Berggren completeness then reads
\begin{eqnarray}
\sum_{i}|u_i\rangle\langle \tilde{u_i}|+\int_{L^{+}}dk|u(k)\rangle\langle \tilde{u(k)}|= \hat{{1}}, 
\end{eqnarray}
where $|u_i\rangle$ are discrete states corresponding to bound and
resonant solutions of the single-particle potential, and
$|u(k)\rangle$ are complex-energy scattering states along the
complex-contour $L^+$. In practice, the integral along the complex
continuum is discretized yielding a finite discrete basis set.

\section{Results}

\label{secnumerics}
We present here applications of the coupled-cluster Green's function approach for \rm{$^{40}$}Ca and \rm{$^{48}$}Ca.
We show results for the bound states in $^{41}$Ca and $^{49}$Ca as well as for the neutron elastic scattering.

Calculations are performed using the $\rm {NNLO_{sat}}$ chiral interaction \cite{n2losat}
which reproduces the binding energy and charge radius for both systems~\cite{hagen2015,garciaruiz2016}.
All results reported here are obtained from coupled-cluster
calculations truncated at the CCSD level, while the PA-EOM and PR-EOM
Lanczos vectors have been truncated at the 2p-1h and 1p-2h excitation level, respectively. 

We first perform HF calculations in a single-particle basis that
employs a mixed representation of harmonic oscillator and Berggren
states. More precisely, to calculate a neutron-target optical potential in the  $(l,j)$ channel, 
we use only Berggren states for the $(l, j)$ neutron partial wave, whereas the rest are taken as harmonic oscillator shells.
We include all harmonic oscillator shells such that $ {2n + l  \leq N_{max}}$. We checked that the results do not
require a special treatment of the continuum in the other partial waves.
The Berggren states are introduced as a discretized set of $N_{berg}=50$ states along a contour in the
complex-k plane up to $k_{max}=4 fm^{-1}$.  This is sufficiently precise to ensure that results are
independent of the form of the contour in the complex-plane.

The  $\rm {NNLO_{sat}}$ interaction includes a two-body and three-body terms. Let us denote $N_2$ and $N_3$ the cutoffs of the interaction terms 
 defined respectively as the maximum number of
quanta allowed in the relative motion of two nucleons and
three nucleons. In all calculations here, we always take $N_2=N_{max}$ and $N_3$ is taken equal to $N_{max}$, except for
the most extensive calculations considered here where $N_{max}=14$ and $N_3=16$.
Moreover, the three-nucleon forces are truncated at the normal-ordered two-body level in the HF basis \cite{ccemerg}.
 The harmonic oscillator frequency is kept fixed at $\hbar \omega$ = 16 MeV. 

To provide perspective, we  also show results for the neutron elastic scattering obtained with the chiral $NN$
interaction $\rm {NNLO_{opt}}$ interaction~\cite{n2loopt}. In that case, the calculations were carried out for $N_{max}=14$ with the harmonic oscillator frequency $\hbar \omega$=20 MeV.

We begin by studying the numerical convergence for the bound states and the associated optical potentials
in $\rm{^{41}Ca}$ and $\rm{^{49}Ca}$.
\subsection{Convergence for bound states} 
\label{sec-conv1} 
The energy for the bound states in $\rm{^{41}}$Ca and $\rm{^{49}}$Ca,
solutions of the PA-EOM CCSD equations, are shown in Tab. \ref{tab-bs}
as a function of $N_{max}$.  For both nuclei, there are only
three bound states supported by the $\rm{NNLO_{sat}}$ Hamiltonian. As
expected, the convergence pattern is slower for the higher-energy
states. Specifically, for $\rm{^{41}Ca}$, the difference between the energies obtained for
$(N_{max},N_3)$=(14,14) and (14,16) is $\sim$ 220 keV in the case of the
ground-state, whereas it is $\sim$350 keV in the case of the $J^{\pi}=1/2^-$ second
excited state. 
For $\rm{^{49}Ca}$, the difference is $\sim$210 keV for the ground state and
$\sim$420 keV for the $J^{\pi}=5/2^-$ excited state.

Even though the absolute binding energy is underestimated in the CCSD approximation, when compared to experiment
(for $\rm{^{40}Ca}$ we obtain a binding energy of 299.28 MeV for
$(N_{max},N_3)=(14,16)$, whereas the experimental value is
342.05 MeV), the neutron separation energies are consistently within
600 keV  of the experimental values.
By including both perturbative triples excitations and perturbative estimates for the neglected
residual 3NFs (3NF terms beyond the normal-ordered two-body
approximation), a good agreement with experimental binding
energies can be obtained for $^{40,48}$Ca \cite{hagen2015}.
\begin{table}[t!]
\begin{ruledtabular}
\begin{tabular}{l l| r r r r}
 & ${N_{max}}$  & ${E(7/2^{-})  }$ & $ { E(3/2^{-}) }$ & ${E(1/2^{-}) }$ \\
\hline 
$\rm^{41}Ca$ & 	& 				& 				&			&		 \\
& 12  & -7.35 & -3.47 & -1.31    \\
& 14 & -7.62 & -3.87 & -1.80   \\
& 14 ($N_3$=16) & -7.84 & -4.07 & -2.15   \\
Exp &       & -8.36	& -6.42	& -4.74	 \\ 
\hline 
$\rm^{49}Ca$ & 					&	$ E( 3/2^{-}) $			&	$E (1/2^{-})$		& $E (5/2^{-})$
\\
&                     12 & -3.88 & -2.025 & -0.37    \\
&                  14 & -4.35 & -2.40  & -1.00   \\
&              14 ($N_3$=16)  & -4.56 & -2.45  & -1.42   \\
Exp &       & -5.14	& -3.12	& -1.56	 \\ 
\end{tabular}
\end{ruledtabular}
\caption{PA-EOM CCSD  energies (in MeV) for bound states in ${\rm ^{41}Ca}$ and ${\rm ^{49}Ca}$ calculated with the chiral ${\rm
 NNLO_{sat}}$ interaction as a function of $N_{max}$. }
\label{tab-bs}
\end{table}

\begin{figure}[htb]
\begin{center}
\includegraphics[scale=0.3]{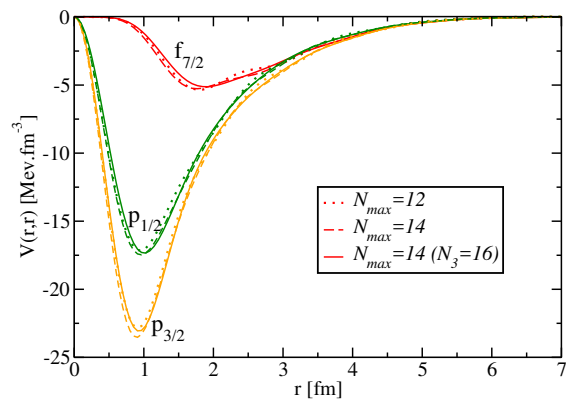}
\caption{ (Color online) Real part of the radial (diagonal) optical potential for the
  bound states in $\rm { ^{41}Ca}$, calculated with the ${\rm  NNLO_{sat}}$ interaction.  Results are shown for 
  the $f_{7/2}$, $p_{3/2}$ and $p_{1/2}$ neutron partial waves and for several
  values of $N_{max}$. }
\label{fig:bound41ca}
\end{center}
\end{figure}

We show for illustration in Fig.~\ref{fig:bound41ca}, the converging pattern of the  real part
of the radial (diagonal) optical potential for the three bound states
in $^{41}$Ca. By construction, the calculated eigenenergies of these potentials are equal to 
 the bound states energies in Tab.~\ref{tab-bs}  when using the effective
mass $m \; A/(A-1)$ instead of the actual reduced mass $m \; (A-1)/A$.  This can be traced to
 Eq.~(\ref{hami2}) where the effective mass associated with
the kinetic operator is equal to $m \; A/(A-1)$.

In the following, we study the convergence pattern for the neutron elastic scattering on $\rm{^{40}}$Ca and $\rm{^{48}}$Ca and
the corresponding optical potentials.
\subsection{Convergence for scattering states} 
\label{sec-conv2} 
We now turn our attention to the neutron elastic scattering on $^{40}$Ca and $^{48}$Ca.  For each partial wave, the scattering phase shift is calculated from the single-particle Schr\"odinger equation using the  optical potential in Eq.~(\ref{optdef}) and the reduced mass  $m \;(A-1)/A $.
Few comments are in order here. Since the calculations  of the optical potential are
performed using the laboratory coordinates (the Hamiltonian H in Eq.~(\ref{hami2}) is defined with these coordinates), the
optical potential in  Eq.~(\ref{optdef}) is obtained in these coordinates and
not in the relative  neutron-target coordinate. However, we will assume here that we can identify the potential in the
relative coordinate with the potential  calculated with
Eq.~(\ref{optdef}). The error associated with this approximation gets smaller as the mass of the nuclei involved increases \cite{rcjohn}.
 
We show in Tab.~\ref{tabphase} the real part of the phase shifts for a
few partial waves for each isotope, at a given scattering beam energy.
The results are shown as a function of $N_{max}$ for neutron scattering at E=5.17~MeV on $\rm{^{40}}$Ca, and at E=7.81~MeV on   $\rm{^{48}}$Ca.
One can see that some of the phase shifts in  Tab.~\ref{tabphase}  are well converged whereas  there are
variations with the model-space sizes in other cases. 
For instance,  convergence is reached for all
but the $s_{1/2}$ and $d_{3/2}$ partial waves for the neutron scattering on $^{40}$Ca at 5.17 MeV.
For the neutron scattering on $^{48}$Ca at 7.81 MeV, the phase shifts are converged for all but the
$g_{7/2}$ partial wave. This difference in the convergence pattern is not unexpected and 
we have found that the partial waves for which
the phase shifts converge slower correspond to those that exhibit a
stronger energy dependence around the energy of interest.

For both systems, the phase shifts should have a finite imaginary part which reflects the loss of flux in the elastic channel.
For instance, in the case of $\rm{^{40}}$Ca at E=5.17~MeV, there is a potential absorption due to excitation of $\rm{^{40}Ca}$
to either its first excited state  $\rm{E(0^+)= 3.35}$ MeV  or second excited state  $\rm{E(3^-)= 3.74}$ MeV.
However, calculations yield  a negligible value for the absorption in all partial waves.
We will return to this point in the next section.
\begin{table}[t!]
\begin{ruledtabular}
\begin{tabular}{l l | r r r r r}
 & $ {N_{max}}$ &  $\delta_{s_{1/2}} $ & $\delta_{p_{1/2}} $ & $\delta_{p_{3/2}} $   & $\delta_{d_{3/2}} $  & $\delta_{d_{5/2}} $   \\
\hline
$\rm^{40}Ca$  	  &	(E=5.17 MeV)			& 				&			&	&	 \\
                        &	 12   & -93 & 80 & 89  & 1 & -99    \\
                         &       14 & -75 & 77 & 88  & 18 & -86 \\
                        &   14 ($N_3$=16)  & -68 & 78 & 89  & 35 & -86 \\
\hline
	&		  & $\delta_{s_{1/2}} $ & $\delta_{p_{1/2}} $ & $\delta_{p_{3/2}} $   & $\delta_{g_{7/2}} $  & $\delta_{g_{9/2}} $   \\
\hline
$\rm^{48}Ca$    & (E=7.81 MeV)	& 				& 				&			&	&	 \\
                           &     12 & -93  & 55  & 62   & 13  & -10     \\
                           &     14 & -83 & 53 & 65 & 22  & -10  \\
                           & 14 ($N_3$=16) & -85 & 53 & 69 & 37 &  -11 \\
\end{tabular}
\end{ruledtabular}
\caption{Real part (in degrees) of the neutron scattering phase shifts calculated with the ${\rm NNLO_{sat}}$
  Hamiltonian at 5.17 MeV  for $\rm^{40}Ca$ and, 7.81 MeV for
  $\rm^{48}Ca$. Results are shown as a function of $N_{max}$.}
  \label{tabphase}
\end{table}
\begin{figure}[t]
\begin{center}
\includegraphics[scale=0.3]{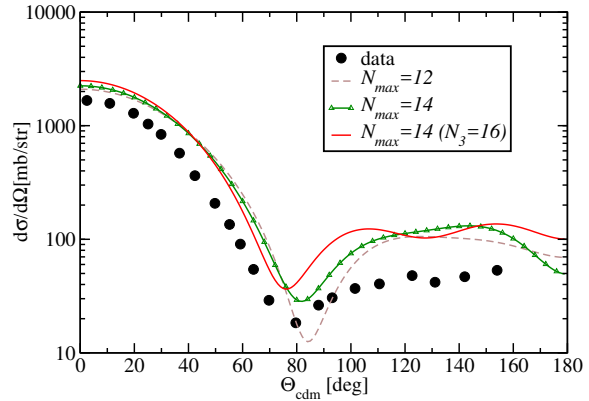}
\caption{ (Color online) Differential elastic cross section for $^{40}$Ca(n,n)$^{40}$Ca at 5.17 MeV, calculated with the ${\rm NNLO_{sat}}$ interaction,  as a function of $N_{max}$. Data points are taken from \cite {KD}.}
\label{fig:xsconv}
\end{center}
\end{figure}
\begin{figure}[t]
\begin{center}
\includegraphics[scale=0.3]{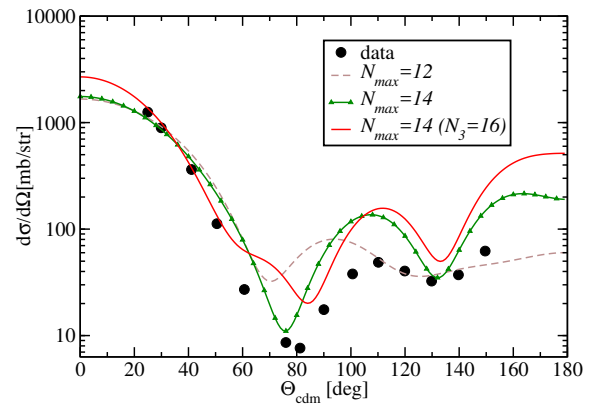}
\caption{ (Color online) Differential elastic cross section for $^{48}$Ca(n,n)$^{48}$Ca at 7.81 MeV, with the ${\rm NNLO_{sat}}$ interaction,
  as  a function of $N_{max}$. Data points are taken from \cite {KD}.}
\label{fig:cont_48}
\end{center}
\end{figure}

Next, we obtain elastic angular distributions by summing the contribution from each partial wave.
We show in Figs.~\ref{fig:xsconv} and \ref{fig:cont_48} the angular distribution
for $^{40}$Ca(n,n)$^{40}$Ca at 5.17 MeV and $^{48}$Ca(n,n)$^{48}$Ca at
7.81 MeV, as  a function of $N_{max}$. We find that the inclusion of partial
waves with angular momentum $L\leq5$ and $L\leq6$ is sufficient for  $\rm{^{40}Ca}$ and $\rm{^{48}Ca}$, respectively, the contribution of partial waves with higher $L$ being negligible. All scattering phase shifts other
than the ones shown in Tab.~\ref{tabphase} have converged with respect to $N_{max}$. 
The variations around the first minimum are significant and are a consequence of the convergence pattern of the scattering phase shifts
with $N_{max}$. We understand then that the calculated cross sections will contain an error due to the model-space truncation.
Note that however, for  $(N_{max},N_3)=(14,16)$, the calculated distribution for $^{48}$Ca(n,n)$^{48}$Ca
is already in excellent agreement with the data at lower angles where
the differential cross section is the largest.
\begin{figure}[htb]
\begin{center}
\includegraphics[scale=0.3]{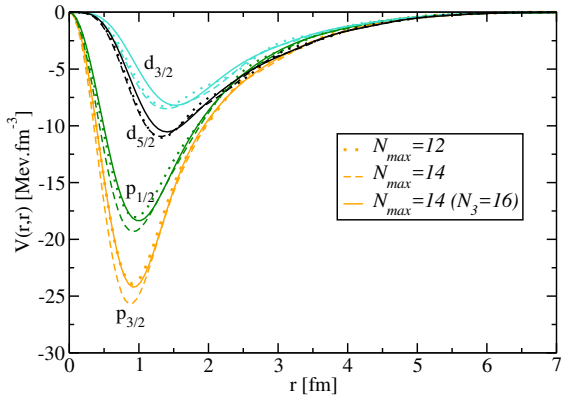}
\caption{(Color online) Real part of the diagonal optical potential for $^{40}$Ca(n,n)$^{40}$Ca at E=5.17 MeV, calculated with the ${\rm NNLO_{sat}}$ interaction, as a function of  $N_{max}$.  Results are shown for the neutron $p_{1/2}, p_{3/2}, d_{3/2}, d_{5/2}$ partial waves.}
\label{fig:scatt41ca}
\end{center}
\end{figure}

Finally, we show for illustration in Fig.~\ref{fig:scatt41ca}, the converging pattern of the real
(diagonal) part of the optical potentials in the neutron $p,d$ partial waves for
$^{40}$Ca(n,n)$^{40}$Ca at 5.17 MeV as a function of $N_{max}$. All corresponding phase shifts  
have converged excepted  for the  $d_{3/2}$ partial wave (cf. Tab.~\ref{tabphase}).
\subsection{Results with finite values of $\eta$} 
\label{sec-eta} 
As we mentioned previously, the calculated optical potentials for neutron scattering
on $\rm{^{40}Ca}$ at E=5.17 MeV and $\rm{^{48}Ca}$ at 7.81 MeV both have negligible absorption.

In order to understand this feature better, let us  consider in more  details the scattering on $\rm{^{40}Ca}$ at 5.17~MeV.
In that case, there is enough energy for the scattered neutron  
to excite the target ($\rm{^{40}Ca}$ in its ground state)
to its two first excited states located at   $\rm{E(0^+)= 3.35}$ MeV  and $\rm{E(3^-)= 3.74}$ MeV.
However the  $0^+$ excited state (which is known to have significant $4p-4h$ components) is
not properly captured in the EOM-CCSD approximation: its calculated energy is at 15.98~MeV above the ground state.
On the other hand, the energy of the second excited state is well reproduced at the EOM-CCSD level with $\rm{E_{EOM-CCSD}(3^{-})=}$3.94 MeV. 
Consequently, only potential excitation to the second excited state could  in principle be accounted for.
The fact that the calculated absorption is nevertheless negligible implies that
the CCSD and PA-EOM wave functions are not sufficiently correlated. In other words,  correlations beyond the singles and doubles truncation
level are needed  to account for the absorption due to target excitation.
The situation is similar for the scattering off  of $\rm{^{48}Ca}$ : in that case, the position of the
first excited state  $\rm{E(2^+)= 3.83}$ MeV is fairly well reproduced at the EOM-CCSD level with  $\rm{E_{EOM-CCSD}(2^{+})=}$4.65 MeV, but the
calculated absorption is still negligible pointing out again to a lack of correlations in the  CCSD and PA-EOM  wavefunctions. 

We should also note that the formation of a compound nucleus will contribute to
 flux removal from the elastic channel. However, again, at that level of truncation, this cannot
 be accounted for since the compound states  consist of a high number of
 particle-hole excitations and are usually described by stochastic
approaches \cite{mitchell2010}.

We  have seen here that one would need to consider excitations beyond the singles and doubles excitations to describe the absorption seen in nature.
However, a cheaper solution may be provided by artificially  considering finite values of
$\eta$ instead of taking the limit $\eta \rightarrow 0^+$ (see Eq.~\ref{gfcc}).
In the following, we explore the impact of using finite $\eta$ values on the optical potential and the scattering phase shift.

\begin{figure}[htb]
\begin{center}
\includegraphics[scale=0.4]{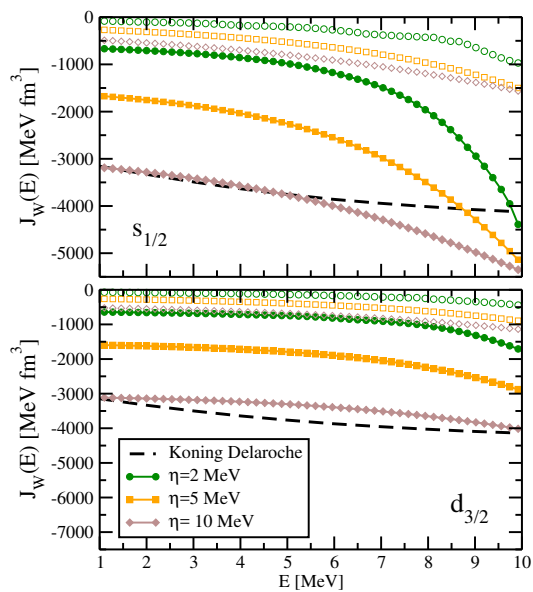}
\caption{(Color online) Volume integral $J_W(E)$ of the imaginary part of the neutron optical potential in the $s_{1/2}$  (upper panel) and
   $d_{3/2}$  (lower panel) partial wave  for $^{40}$Ca(n,n)$^{40}$Ca as a function of energy E. Results  are shown for the $\rm{NNLO_{sat}}$
  (symbols with lines) and $\rm{NNLO_{opt}}$ (symbols without lines) interactions at $\eta=2,5,10$ MeV.
  Results for the KD potential are also shown for comparison.}
\label{fig:J_im}
\end{center}
\end{figure}
\begin{figure}[htb]
  \begin{center}
    \includegraphics[scale=0.3]{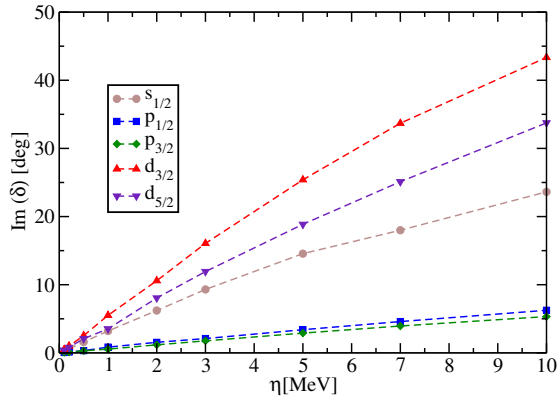}
    \caption{(Color online) Imaginary part of the phase shift in
      the $spd$ partial waves for $^{40}$Ca(n,n)$^{40}$Ca at 5.17 MeV as a function of $\eta$.} 
    %  Results are shown  for $(N_2,N_3)=(14,16)$.
       \label{fig:phase_eps}
  \end{center}
\end{figure}
We show in Fig.~\ref{fig:J_im} the imaginary volume integral $J_W^l(E)$, 
\begin{eqnarray}
J_W^l(E)=4\pi\int dr r^2 \int dr r'^2 {\rm Im} \Sigma'_l(r,r';E)
\end{eqnarray}
of the optical potential for $^{40}$Ca(n,n)$^{40}$Ca in the $s_{1/2}$ and $d_{3/2}$ partial waves with $\eta=2,5,10$ MeV. 
Results are shown for the  $\rm{NNLO_{sat}}$ and $\rm{NNLO_{opt}}$  interactions and the KD potential.
While these quantities are not observables, the comparison with the integral of the KD potential is instructive
(see Fig.~\ref{fig:J_im}) and underscores  the lack of significant absorption of the potential calculated with
the coupled-cluster Green's function at the singles and doubles approximation level.
Obviously, increasing the value of $\eta$,  increases the  value of the integrals (in modulus) and  consequently the neutron absorption in the scattering reaction.
This is further illustrated in Fig.~\ref{fig:phase_eps} where we show the imaginary part of the
scattering phase shifts for $^{40}$Ca(n,n)$^{40}$Ca at 5.17 MeV as a function of $\eta$.
For $\eta=$0 MeV, all phase shifts have a vanishing imaginary part and  as $\eta$ increases,
the imaginary parts increase more or less depending on the partial wave considered. 
If we were interested in reproducing the volume integral of the KD potential at 5.17 MeV in the $s_{1/2}$  partial wave, one
would choose a value of $\eta\sim 10$ MeV (see  Fig.~\ref{fig:J_im}).

In the following section, we show results for the elastic cross section on $\rm{^{40}Ca}$  and  $\rm{^{48}Ca}$ 
with increasing absorption by using finite values of $\eta$.
\subsection{Results for elastic scattering}
\label{sec-results}
We now discuss predictions for the elastic cross section when
considering values of  $\eta=0, 2, 5$ MeV, for $^{40}$Ca and
$^{48}$Ca.   All calculations presented
in this section correspond to the largest model space discussed in the
previous section namely $N_{max}=14$ and $N_3=16$.

\begin{figure}[h]
\begin{center}
\includegraphics[scale=0.4]{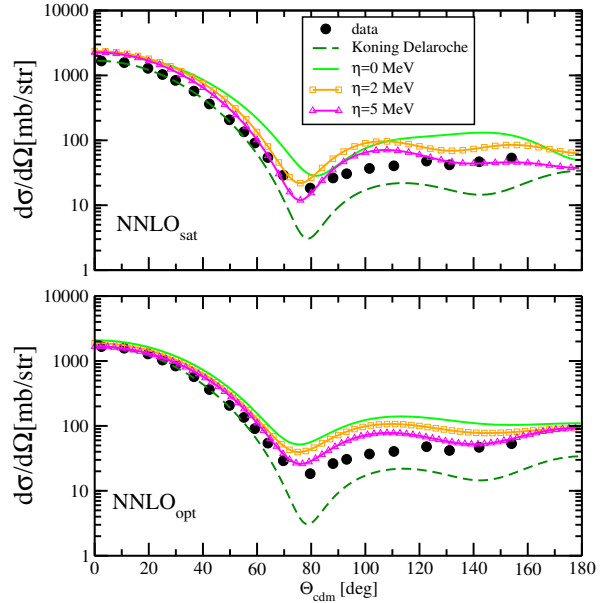}
\caption{(Color online) Differential elastic cross section for $^{40}$Ca(n,n)$^{40}$Ca at  5.17 MeV calculated with the $\rm{NNLO_{sat}}$
 (top)  and $\rm{NNLO_{opt}}$ (bottom) interactions. Calculations are shown for  $\eta=0,2,5$ MeV.
  Results obtained using the KD  potential are shown for comparison. Data points are taken from \cite {KD}.}
\label{fig:cont4_40}
\end{center}
\end{figure}

\begin{figure}[htb]
  \begin{center}
\includegraphics[scale=0.4]{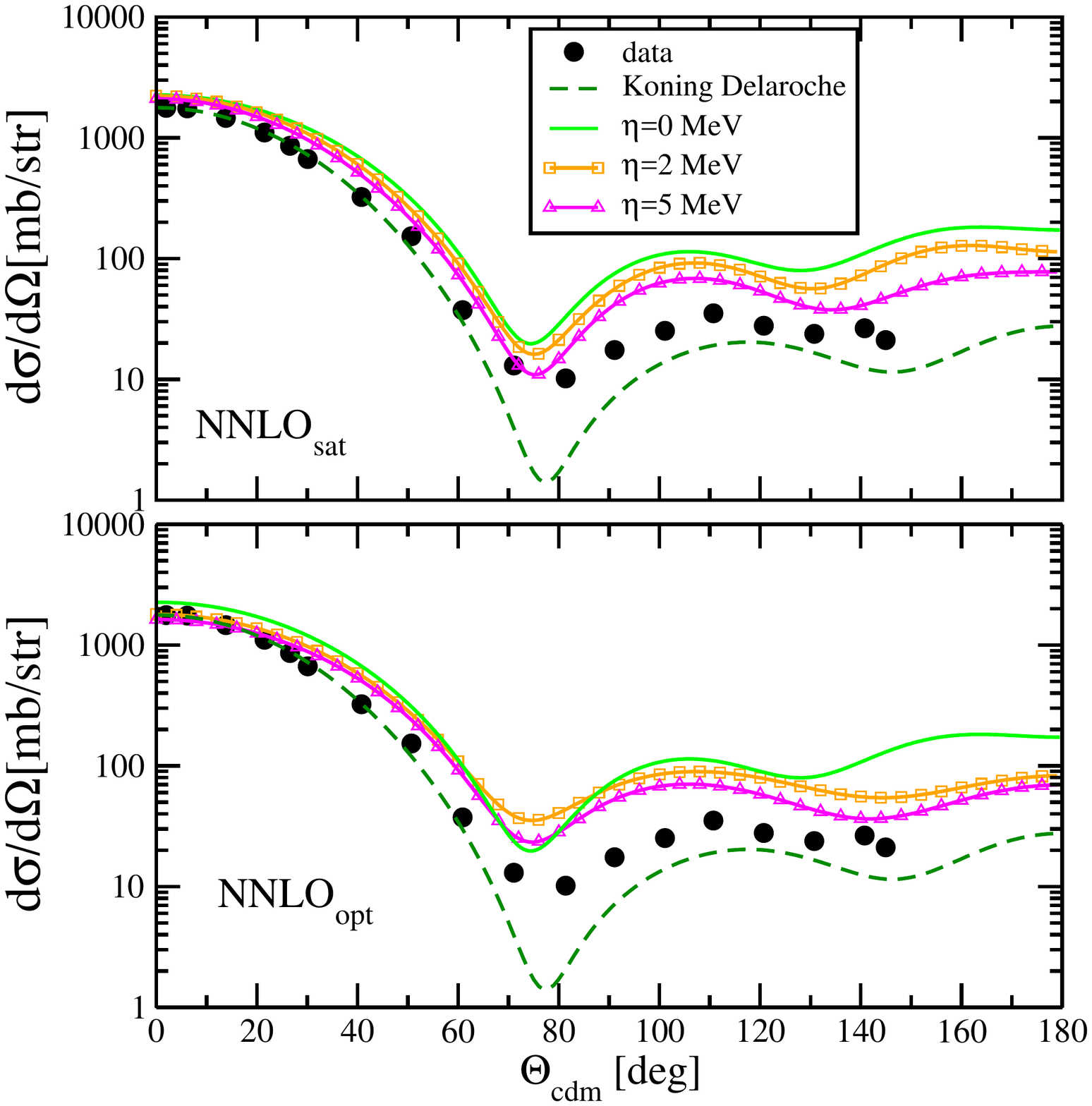}
\caption{(Color online) Differential elastic cross section for $^{40}$Ca(n,n)$^{40}$Ca at  6.34  MeV calculated with the $\rm{NNLO_{sat}}$ (top)
  and $\rm{NNLO_{opt}}$ (bottom) interactions. Calculations are shown for  $\eta=0,2,5$ MeV.
   Results obtained
using the KD  potential are shown for comparison. Data points are taken from \cite {KD}.}
  \label{fig:plot_6_5_sat_eps}
\end{center}
\end{figure}

The calculated differential elastic cross sections for neutron
scattering on $^{40}$Ca at E=5.17 MeV and E=6.4 MeV are shown in
Fig.~\ref{fig:cont4_40} and Fig.~\ref{fig:plot_6_5_sat_eps}
respectively. The top (bottom) panel corresponds to the results using
the $\rm{NNLO_{sat}}$ ($\rm{NNLO_{opt}}$) interaction. For comparison, we
also show the angular distributions obtained with the phenomenological
KD potential, and also the measured cross sections (errors on the data
are smaller than the symbols).  As expected, when $\eta$ increases, the
elastic scattering cross section decreases with a more pronounced
(relative) reduction at larger angles. Moreover, the agreement with data improves as $\eta$ increases.
 The level of  disagreement between the experimental data and the result obtained
with KD is an illustration of the level of accuracy that can be
expected from a phenomenological global interaction.

\begin{figure}[h]
\begin{center}
\includegraphics[scale=0.4]{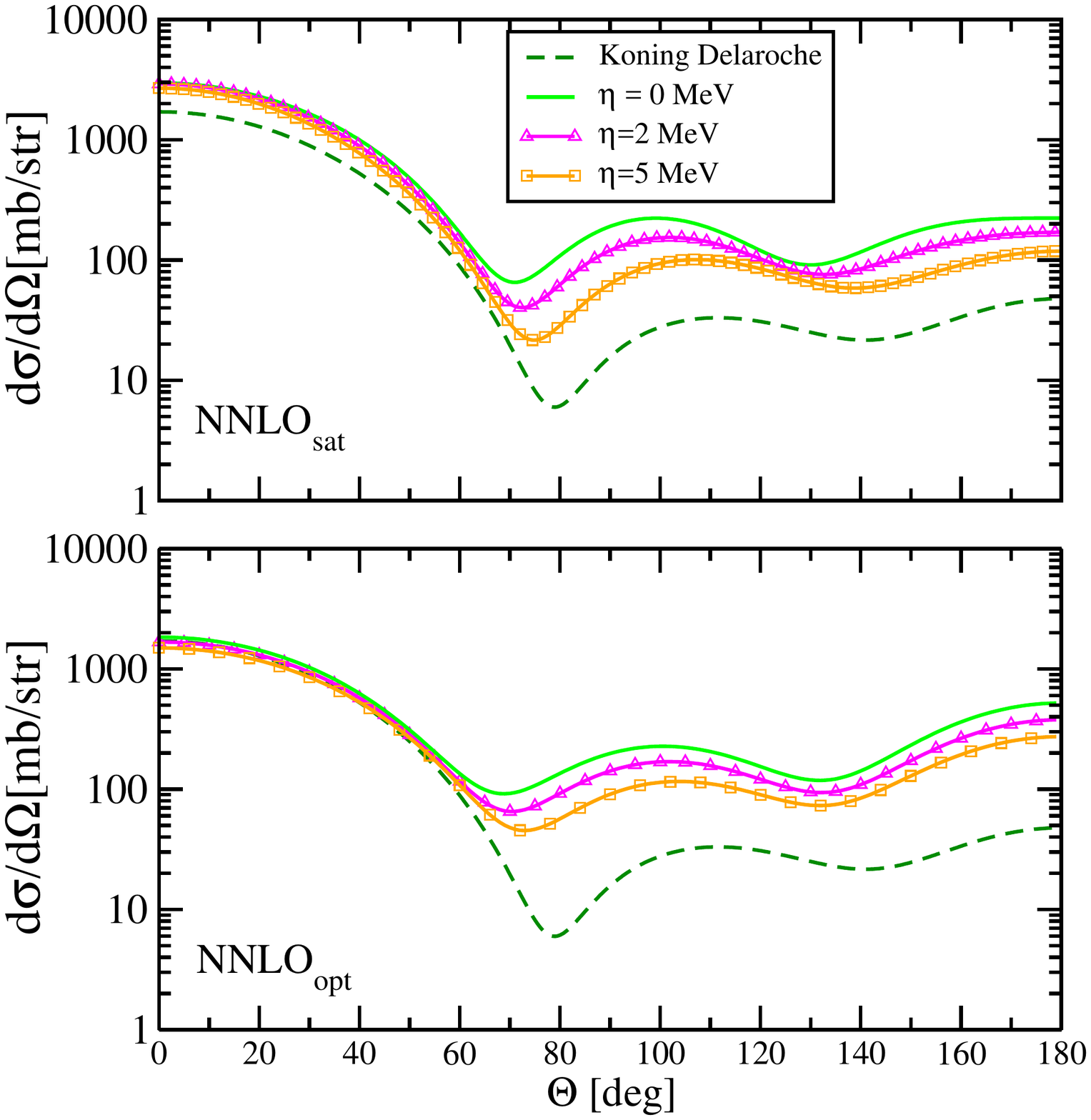}
\caption{(Color online) Differential elastic cross section for $^{48}$Ca(n,n)$^{48}$Ca at 4.00 MeV calculated with the $\rm{NNLO_{sat}}$ (top)
  and $\rm{NNLO_{opt}}$ (bottom) interactions. Calculations are shown for  $\eta=0,2,5$ MeV.
  Results obtained
using the KD  potential are shown for comparison. 
}
\label{fig_48_el1}
\end{center}
\end{figure}
\begin{figure}[h]
\begin{center}
\includegraphics[scale=0.4]{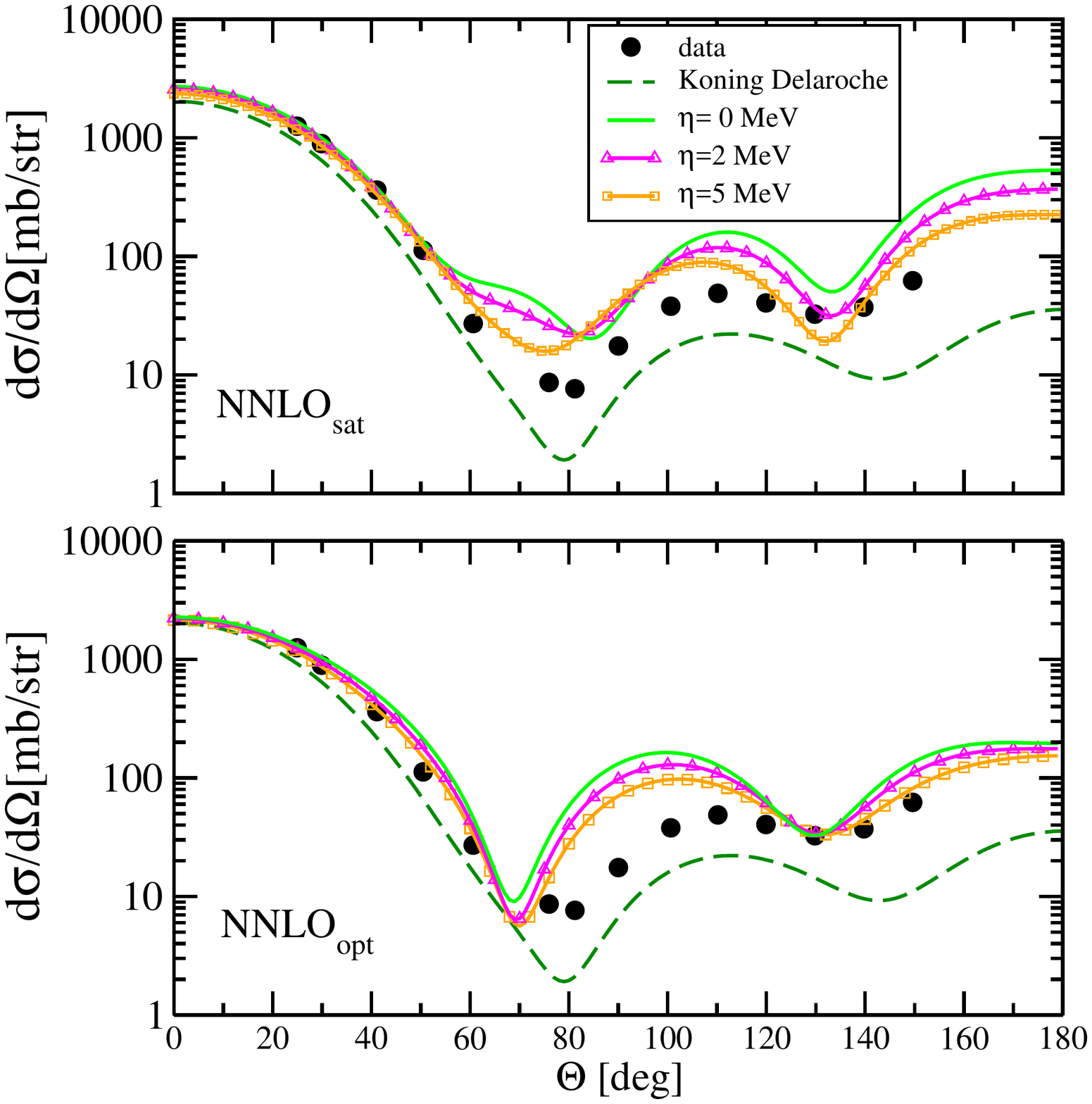}
\caption{(Color online) Differential elastic cross section for $^{48}$Ca(n,n)$^{48}$Ca at 7.81 MeV calculated with the $\rm{NNLO_{sat}}$ (top)
  and $\rm{NNLO_{opt}}$ (bottom) interactions. Calculations are shown for  $\eta=0,2,5$ MeV.
    Results obtained
using the KD  potential are shown for comparison.
Data points are taken from \cite {KD}.}
\label{fig_48_el2}
\end{center}
\end{figure}

Next, we show predictions for neutron elastic scattering on $^{48}$Ca at E=4 MeV (Fig.~\ref{fig_48_el1}) and E=7.81
MeV (Fig.~\ref{fig_48_el2}). There is no data available for the neutron elastic scattering at 4 MeV,
but we chose to include it in our study to show that
the general behavior of increasing absorption is the same
independently of the scattering energy and the target system.  As  for  
for $^{40}$Ca, it is also clear that when including $\eta$ as a fine-tuning
parameter we can improve the agreement with data. Note that  even at $\eta=0$ MeV, the calculated distribution
 for  $\rm{NNLO_{sat}}$ in Fig.~\ref{fig_48_el2} is in excellent agreement with the data at smaller angles where
the differential cross section is the largest.
The same characteristics shown for
$^{40}$Ca are present in $^{48}$Ca case, namely the cross section is not strongly
dependent on $\eta$ at small angles, while at larger angles the cross
section is significantly reduced with increasing $\eta$.
Moreover, for both nuclei, the results show a sensitivity of the distributions to the
employed Hamiltonian.

An encouraging result of our calculations is that, within the energy
range considered in this work, fine-tuning $\eta$ allows us to improve
the description of neutron elastic scattering for both $^{40}$Ca and
$^{48}$Ca. The value of $\eta$ we use should not be interpreted as the
effective width of the states, but rather as a means to compensate for
the truncations inherent to our approach.
\begin{table}[htb]
\begin{ruledtabular}
\begin{tabular}{l | r |  r r r | r}
A		&	E (MeV) & $\eta=0$ MeV & $\eta=2$ MeV & $\eta=5$ MeV  & KD  \\[2pt]
\hline
40  & 5.17 & 229(12)	 &   195(13)  & 166(12)   & 108 \\
 & 6.3 	&  195(3) &    169(10) &  144(9)  & 96 \\
48  & 7.81 & 182 (32) &   159(13) & 139 (12)  & 88 \\
\end{tabular}
\end{ruledtabular}
\caption{Total elastic  cross sections (in b) calculated with the ${\rm NNLO_{sat}}$ interaction for $^{40}$Ca and $^{48}$Ca.
 Results are shown for $\eta=0,2,5$ MeV. For each case, we  assign an error defined as
the difference between the calculated cross section obtained with $(N_{max},N_3)=(14,16)$ and $(N_{max},N_3)=(14,14)$.
Results obtained with the KD potential are  shown for comparison.}
  \label{tab_xstot}
\end{table}

Finally in Tab.~\ref{tab_xstot}, we show the total elastic cross
sections for both isotopes and energies, as a function of $\eta$.
These were obtained by integrating the differential cross sections
over angle. We only show the value for ${\rm NNLO_{sat}}$ (the same
features appear when using ${\rm NNLO_{opt}}$). We also include an
error based on the model space truncation: we assign an error as being
the difference between the total cross section obtained with
$(N_{max},N_3=(14,16)$ and that for $(N_{max},N_3)=(14,16)$.  In the last
column, we show the results using the phenomenological interaction
KD. In all cases, we can see that the cross sections calculated with the
coupled-cluster optical potential are larger than the predictions obtained
with the KD potential.  

\section{Conclusions} \label{conclusion} In this paper we constructed
microscopic nuclear optical potentials by combining the Green's
function approach with the coupled-cluster method at the singles and doubles truncation level. We used an
analytical continuation in the complex-energy plane, based on a
complex Berggren basis, to compute the Green's function. The Dyson
equation was then inverted to obtain the optical potential. We showed
applications for  $^{40}$Ca and $^{48}$Ca with the chiral $NN$ and 3NF
interaction $\rm{NNLO_{sat}}$. The choice of this interaction was
motivated by the fact that it allows for a good description of masses
and radii in a wide mass-range and, furthermore, it reproduces the
charge radii of $\rm{^{40}Ca}$ and $\rm{^{48}Ca}$. First we showed
results for the optical potentials associated with the bound states in $^{41}$Ca and $^{49}$Ca and then
presented applications to the neutron scattering. We 
also showed, for comparison, the results for  neutron scattering obtained with the chiral $NN$ interaction
$\rm{NNLO_{opt}}$, and with the phenomenological Koning-Delaroche
potential. We have seen that the overall form of the scattering cross
section is reproduced for both nuclei at several scattering
energies. At this level of truncation, the absorption is practically
negligible which points to a lack of many-body correlations in the wave
functions of the coupled-cluster method at the singles and doubles
approximation level. We showed that, by increasing the parameter $\eta$
in the Green's function, results can be somewhat improved.

This work can be extended in several directions. We plan to consider
higher-order correlations in our coupled-cluster Green's function
calculations as was recently done for the dipole response of
$^{48}$Ca~\cite{miorelli2018}, and excited states in
$^{101}$Sn~\cite{morris2018}. The first step will be to include
iterative triples excitations in the ground-state, and investigate
the impact of these correlations on the absorptive character of the
calculated optical potential. It could also be interesting to investigate optical 
potentials constructed by starting with the singles and doubles coupled-cluster Green's function potential and add an {\it ad hoc}  polarization terms which would effectively account
for the  missing physics (such as collective excitations and formation of compound nucleus)  at that level of truncations.

\begin{acknowledgments}
We thank K. Hebeler for providing us with matrix elements in Jacobi
  coordinates for the $NNN$ interaction at next-to-next-to-leading order.
  We acknowledge beneficial discussions with Willem Dickhoff, Charlotte
 Elster, Heiko Hergert and Gr\'egory Potel.  
This work was supported by the National Science Foundation under Grant  PHY-1403906,  the Department of Energy under Contract No. DE-FG52-
08NA28552, by the Office of Science, U.S. Department of Energy under Award Number DE-SC0013365  
and by the Office of Nuclear Physics,
U.S. Department of Energy, under Grants DE-SC0008499 (SciDAC-3 NUCLEI), DE-SC0018223 (SciDAC-4  NUCLEI), the Field Work Proposal
ERKBP57 and ERKBP72 at Oak Ridge National Laboratory (ORNL).
An award of computer
time was provided by the Institute for Cyber-Enabled Research
at Michigan State University and part of this  research used resources 
of the Oak Ridge Leadership Computing Facility located
at ORNL, which is supported by the Office of Science
of the Department of Energy under Contract No. DE-
AC05-00OR22725.
\end{acknowledgments}

\bibliography{refs}

\end{document}